\documentclass[10pt, twocolumn, twoside]{IEEEtran}
\usepackage{graphicx}
\usepackage{amsmath}
\usepackage{amssymb}
\usepackage{bm}
\usepackage[caption=false,font=footnotesize]{subfig}
\usepackage{cite}

\newtheorem{theorem}{Theorem}
\newtheorem{corollary}[theorem]{Corollary}

\def\argmin{\mathop{\mathrm{arg\,min}}}
\def\MID{\, | \,}
\def\Hhat{\widehat{H}}
\def\hhat{\widehat{h}}
\def\Belief{q}
\def\Rp{{R_{\rm p}}}
\def\Rs{{R_{\rm s}}}
\def\P{\mathbb{P}}
\newcommand{\LocalI} [1]{{P_{e,#1}^{\rm I}}}
\newcommand{\LocalII}[1]{{P_{e,#1}^{\rm II}}}
\newcommand{\LocalIx}{{P_{e}^{\rm I}}}
\newcommand{\LocalIIx}{{P_{e}^{\rm II}}}
\newcommand{\LocalIxSeq}[1]{{P_{e}^{{\rm I}_{#1}}}}
\newcommand{\LocalIIxSeq}[1]{{P_{e}^{{\rm II}_{#1}}}}
\newcommand{\LocalISeq}[3]{{P_{e,#1_{#3}}^{{\rm I}_{#2}}}}    
\newcommand{\LocalIISeq}[3]{{P_{e,#1_{#3}}^{{\rm II}_{#2}}}}
\newcommand{\GlobalI} {{P_{E}^{\rm I}}}
\newcommand{\GlobalII}{{P_{E}^{\rm II}}}
\newcommand{\DecThres}[3]{{\lambda_{#1_{#3}}^{_{#2}}}}

\begin{document}

\title{Distributed Hypothesis Testing with \\ Social Learning and Symmetric Fusion%
\thanks{This material is based upon work supported by the National Science Foundation under Grant No.~1101147\@.
    The material in this paper was presented in part at the IEEE International Conference on Acoustics, Speech, and Signal Processing, Vancouver, Canada, May 2013.}
}

\author{Joong~Bum~Rhim and Vivek~K~Goyal%
    \thanks{J. B. Rhim (email: jbrhim@mit.edu) is with the Department of Electrical Engineering and Computer Science and the Research Laboratory of Electronics, Massachusetts Institute of Technology.}
    \thanks{V. K. Goyal (email: v.goyal@ieee.org) is with the Department of Electrical and Computer Engineering, Boston University.}
}

\markboth{Distributed Hypothesis Testing with Social Learning and Symmetric Fusion}{Rhim and Goyal}

\maketitle

\begin{abstract}
We study the utility of social learning in a distributed detection model with agents sharing the same goal: a collective decision that optimizes an agreed upon criterion.
We show that social learning is helpful in some cases but is provably futile (and thus essentially a distraction) in other cases.
Specifically, we consider Bayesian binary hypothesis testing performed by a distributed detection and fusion system, where all decision-making agents have binary votes that carry equal weight.
Decision-making agents in the team sequentially make local decisions based on their own private signals and all precedent local decisions.  It is shown that the optimal decision rule is not affected by precedent local decisions when all agents observe conditionally independent and identically distributed private signals.  Perfect Bayesian reasoning will cancel out all effects of social learning.
When the agents observe private signals with different signal-to-noise ratios, social learning is again futile if the team decision is only approved by unanimity.
Otherwise, social learning can strictly improve the team performance.  Furthermore, the order in which agents make their decisions affects the team decision.
\end{abstract}


\section{Introduction}
\label{sec:Introduction}

\begin{figure}
    \centering{
    \subfloat[Centralized detection.]{\includegraphics[width=1.7in]{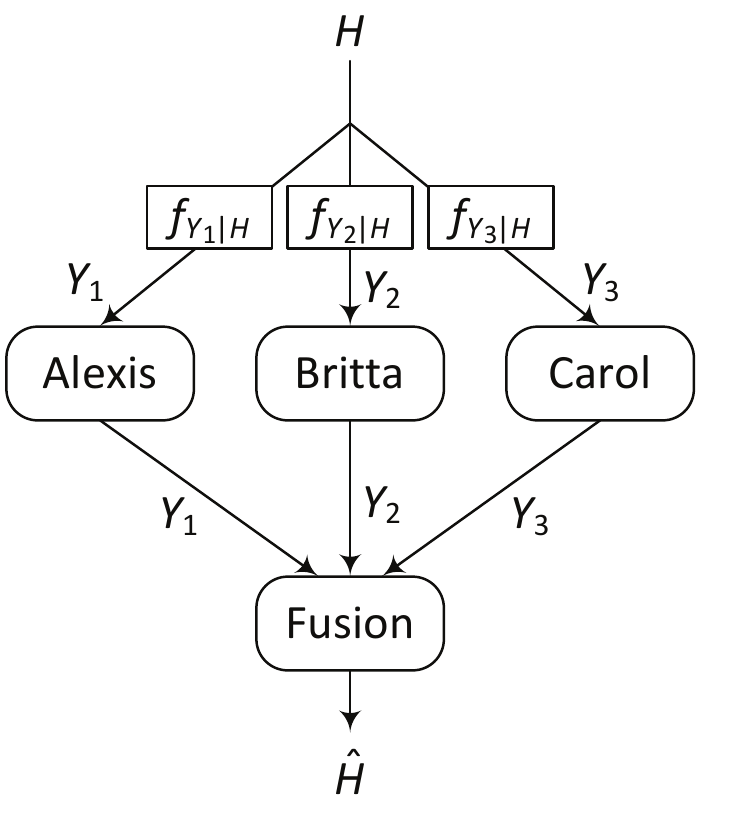} \label{fig:Team1}}
    \subfloat[Conventional distributed detection.]{\includegraphics[width=1.7in]{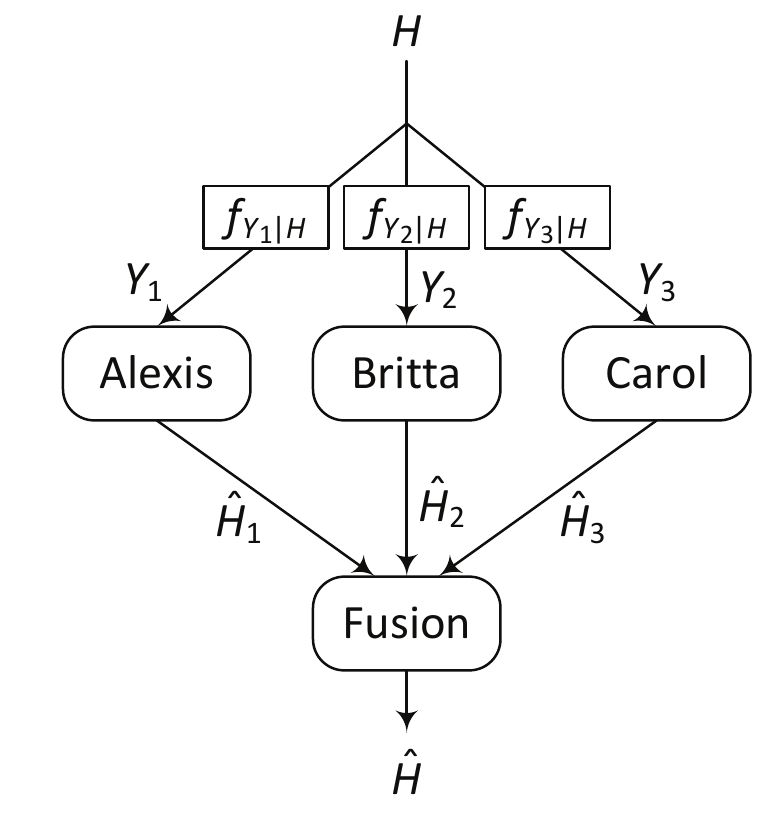} \label{fig:Team2}}
    \hfill
    \subfloat[Focus of this paper.]{\includegraphics[width=2in]{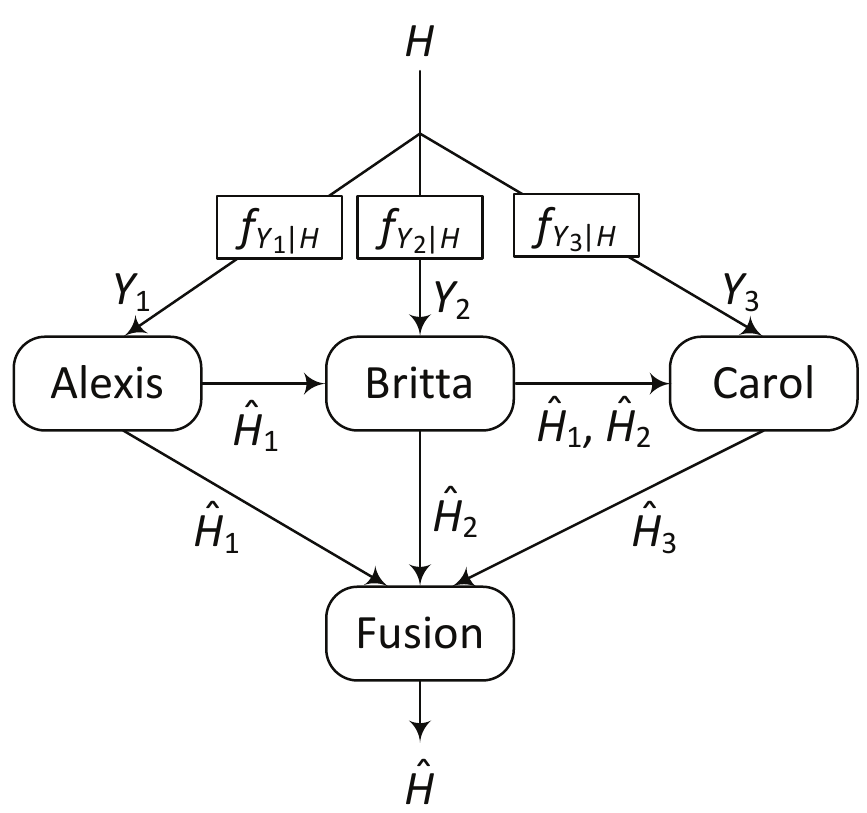} \label{fig:Team3}}
    }
    \caption{Examples of group decision making.  (a) Agents observe private signals and send them to the fusion center for centralized detection.  (b) Agents observe private signals, make local decisions in a distributed matter, and send their decisions to the fusion center.  (c) Agents observe private signals, make local decisions sequentially using all available information, and send their decisions to later-acting agents and to the fusion center.  Sending decisions to later-acting agents is termed \emph{public voting}.}
    \label{fig:Team}
\end{figure}

Consider a team of three physicians, Alexis, Britta, and Carol, who together diagnose a patient and make a decision, say between medication and surgery.  Each physician individually examines the patient or runs medical tests within her expertise before the team aggregates the individual opinions.
The team would make the best decision if the physicians could share every result and discuss how to treat the patient.
This would be as if the agents send their observations in full to a fusion center, which is able to use a sophisticated integration rule, as depicted in Fig.~\ref{fig:Team1}.  However, this level of communication costs a lot of time and resources, and this architecture assumes the ability to jointly interpret the totality of information.

A more common way to aggregate opinions of human decision makers is through some form of voting.
In fact, a patient may seek the opinions of three physicians and decide according to the majority, with no interaction among the physicians, as depicted in Fig.~\ref{fig:Team2}.
This is a typical distributed detection model.  The optimal local decision-making strategy in such model has been studied for decades~\cite{Tsitsiklis1988,Varshney97}.

In some settings, people share their opinions and thus can affect the opinions of others.
In the medical example, the patient may see the physicians in sequence, and the recommendation (medication or surgery) of a physician may be known to each later-consulted physician when she is making her own recommendation.
A simple example is shown in Fig.~\ref{fig:Team3}.
The effect of this interaction among decision-making agents---under certain assumptions that are adopted for tractability---is the focus of this paper.

In our model, agents make decisions in a predetermined order and later-acting agents can observe all or some of the decisions made by earlier-acting agents.  The local decisions are also called \emph{public signals} because they are shown to multiple agents, unlike the \emph{private signals}.  The later-acting agents can perform Bayesian reasoning to learn some information from the public signals and make better local decisions.
This process of incorporating information from other agents is called \emph{social learning}.  We will see that social learning provides no improvement when the agents have conditionally independent and identically distributed (iid) private signals; otherwise there is generally an improvement, and the degree of improvement depends on the order in which the agents make their decisions.

\subsection{Relations to Previous Work}
Mathematical study of social learning was introduced in~\cite{Banerjee1992,BikhchandaniHW1992}.
These papers discussed decision-making agents who make decisions in a predetermined order and for individually maximizing their own utilities.  They presented models where choices made by all agents are observed by all other agents as public signals.  In these models, herding behavior---that agents ignore private signals and adopt public signals---arises after several earlier-acting agents choose the same alternative.
It was later shown that incorrect herding occurs with a nonzero probability if the private signals are boundedly informative.\footnote{A signal $Y$ generated under a state $H$ is called \emph{boundedly informative} if there exists $\kappa > 0$ such that $\kappa < f_{Y | H}(y \MID h) < 1 / \kappa$ for all $y$ and $h$.} Otherwise, asymptotically a sequence of agents will converge to the correct choice~\cite{SmithS2000}.

This work is unique from previous works in that we study social learning in a team decision-making scenario with a symmetric fusion of the local decisions.  The fusion by voting makes an earlier-acting agent important not only because she influences later-acting agents but also because her vote counts in the fusion rule.
We ask whether it is beneficial for the agents to share their local decisions among themselves.
The agents in this work perform Bayesian reasoning, and there is a rich literature that suggests this to be a good approximation to human behavior~\cite{SwetsTB1961,Viscusi1985,BraseCT1998,GillSabinSchmid2005,GlanzerHM2009}.

Note that we consider agents who share a common goal.  The social learning in previous works occurs among agents that have individual goals and behave selfishly to achieve their own goals. Social learning is helpful in this case if the private signals are unbounded~\cite{SmithS2000} even though the selfishness diminishes the efficiency of social learning~\cite{RhimG2013b}.
Chamley \emph{et al}.~\cite{ChamleySL2013} also state that social learning yields a lower rate of learning than collaboration does.  They compare social learning by selfish agents to distributed detection by collaborating agents but do not consider social learning by collaborating agents.  In contrast, we are considering social learning and distributed detection together in one model.

When the agents' private signals are not conditionally iid, we find that the order in which the agents make their decisions can affect the collective performance.  The ordering that maximizes the amount of information transferred to the fusion center was studied previously in the case when agents observe binary private signals by Ottaviani and S{\o}renson~\cite{OttavianiSorensen2001}.  They state that for a given group of heterogeneous expert agents, the order in which the agents speak matters.  The same statement holds in our model even though the agents in our model have a different motivation than the agents in the previous model, who pursue only the personal goal of improving their reputations.
The importance of ordering is also pointed out in~\cite{KrishnamurthyPoor2013}.

Our preliminary work~\cite{RhimG2013a} compares private voting to public voting in which all agents observe the full history of previous decisions and the agents observe conditionally iid private signals.  It is shown that public voting does not improve performance.  Intuitively, one might think that social learning helps later-acting agents make better local decisions, leading to improvement of the team decision; however, this intuition is shown to be wrong.

This paper builds upon the preliminary results in~\cite{RhimG2013a}. We extend our interest to more general models.
We allow the likelihoods for the agents' private signals to differ.
Also, agents may observe only some of previous decisions, such as a decision made by the agent just before themselves in a sequence~\cite{Varshney97} or local decisions made by their neighbors in a specific network~\cite{AcemogluDLO2011}.

All discussions in this paper do not require that private signals are equal to the hypothesis corrupted by additive or Gaussian noises.
Agents observe private signals with likelihood functions that are different under state 0 and state 1\@.  All we need is that their likelihood ratios are strictly monotonic, which is a rather natural requirement.

\subsection{Outline}
Section~\ref{sec:Problem} formally describes our decision-making model.  Section~\ref{sec:IdenticalAgents} provides results for agents observing conditionally iid private signals.  Section~\ref{sec:Experts-Crowd} generalizes the model to agents with differing private signal likelihoods and provides examples of helpful social learning.  Section~\ref{sec:Conclusion} summarizes our results and discusses limitations and extensions of our model.

\section{Problem Statement}
\label{sec:Problem}

\begin{figure}
    \centering
    \includegraphics[width=3.4in]{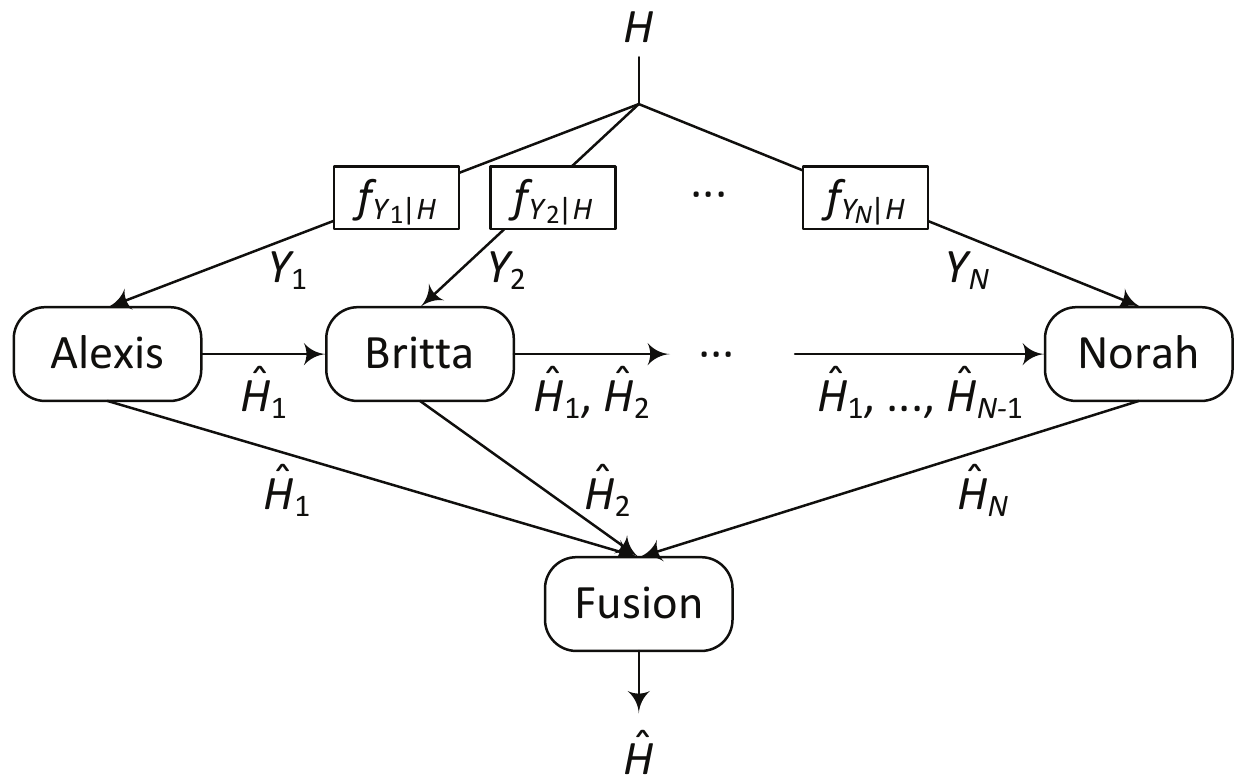}
    \caption{The distributed detection and decision fusion model considered in this paper.  Local decisions are sent not only to the fusion center but also to following agents as public signals.}
    \label{fig:Model_N}
\end{figure}

Consider a distributed detection and decision fusion system with $N$ agents: Alexis, Britta, Carol, Diana, \ldots, Norah.
We also number the agents $1,\,2,\,\ldots,\,N$; names and numbers are used as convenient.
Together, the agents perform a binary hypothesis test to detect $H \in \{0, 1\}$, which has prior probability distribution $p_0 = \P\{H = 0\}$ and $p_1 = \P\{H = 1\} = 1 - p_0$.
Agent $n$ observes a \emph{private signal} $Y_n$ about the true state $H = h$, with likelihood function $f_{Y_n | H}(y_n \MID h)$,
as shown in Fig.~\ref{fig:Model_N}.
The private signals $\{Y_n\}_{n=1}^N$ are conditionally independent given $H$.
We assume that the likelihood ratios $f_{Y_n | H}(y_n \MID 1) / f_{Y_n | H}(y_n \MID 0)$ are monotonically increasing.

In our model, Agent $n$ makes \emph{local decision} $\Hhat_n \in \{0, 1\}$ or, equivalently, quantizes her private signal into 0 or 1, and sends it to a fusion center.  The fusion center uses a fixed and symmetric rule to aggregate the local decisions:
\begin{align}
    \Hhat = \left\{ \begin{array}{@{\,}ll}
                    1, & \mbox{if $\ \sum_{n = 1}^{N} \Hhat_n \geq L$}; \\
                    0, & \mbox{otherwise}, \end{array} \right.
    \label{eq:SymmetricFusionRule}
\end{align}
for some $L \in \{1,\,2,\,\ldots,\,N\}$.
We call this the $L$-out-of-$N$ fusion rule and call the result of the fusion $\Hhat$ a \emph{team decision}.

The goodness criterion for the decision making is the Bayes risk (mean decision-making cost):
\begin{equation}
    R = c_{10} p_0 \P\{\Hhat = 1 \MID H = 0\} + c_{01} p_1 \P\{\Hhat = 0 \MID H = 1\},
    \label{eq:BayesRisk}
\end{equation}
where $c_{10}$ denotes the cost of a false alarm or Type I error---detecting $H = 0$ as $\Hhat = 1$---and $c_{01}$ denotes the cost of a missed detection or Type II error---detecting $H = 1$ as $\Hhat = 0$.  We assume that a correct decision induces no cost.
The agents are a team in the sense of Radner~\cite{Radner1962}:  they share the same decision-making costs and the same goal, which is to minimize~\eqref{eq:BayesRisk}.

Each Bayesian agent performs a likelihood ratio test (LRT) to make optimal local decisions~\cite{NeymanPearson1933,Varshney97}.  The test chooses the hypothesis $\Hhat_n = 1$ if the likelihood ratio is greater than a certain threshold, which is determined by the information the agent has, such as the prior probability, decision-making costs, and the number of agents.
Since we are assuming monotonically increasing likelihood ratios, we can express the LRT in a compact form:
\begin{equation}
    y_n \overset{\Hhat_n(y_n) = 1}{\underset{\Hhat_n(y_n) = 0}{\gtreqless}} \lambda_n,
\end{equation}
where we call $\lambda_n$ a decision threshold.

Each agent generates a local decision $\Hhat_n$ after the LRT\@.  Her decision may be a false alarm or a missed detection, which have probabilities $\LocalI{n}{}{}$ and $\LocalII{n}{}{}$, respectively:
\begin{align}
    \LocalI{n}{}{} & = \P\{\Hhat_n = 1 \MID H = 0\}, \\
    \LocalII{n}{}{} & = \P\{\Hhat_n = 0 \MID H = 1\}.
\end{align}
From the $L$-out-of-$N$ fusion rule, we can compute the probabilities of false alarm and of missed detection of the team decision, $\GlobalI$ and $\GlobalII$:
\begin{align}
    \GlobalI & = \P\left\{\Hhat = 1 \MID H = 0\right\} = \P\left\{\textstyle \sum_{n = 1}^{N} \Hhat_n \geq L \MID H = 0\right\} \nonumber \\
    & = \sum_{n = L}^{N} \sum_{I \subseteq [N] \atop |I| = n} \prod_{i \in I} \LocalI{i} \prod_{j \in [N] \setminus I} \left(1 - \LocalI{j}\right),\nonumber \\
   \GlobalII & = \P\left\{\Hhat = 0 \MID H = 1\right\} = \P\left\{\textstyle \sum_{n = 1}^{N} \Hhat_n < L \MID H = 1\right\} \nonumber \\
   & = \sum_{n = N{-}L{+}1}^{N} \sum_{I \subseteq [N] \atop |I| = n} \prod_{i \in I} \LocalII{i} \prod_{j \in [N] \setminus I} \left(1 - \LocalII{j}\right), \nonumber
\end{align}
where $[N]$ denotes the set $\{1, 2, \ldots, N\}$.  The Bayes risk~\eqref{eq:BayesRisk} can be rewritten as
\begin{equation}
    R = c_{10} p_0 \GlobalI + c_{01} p_1 \GlobalII.
\end{equation}
Note that this is the \emph{team's} Bayes risk,
which is the only goodness criterion considered here;
agents do not minimize a personal Bayes risk.

In a decision-making task, the agents make decisions sequentially in the preordained order of Alexis, Britta, Carol, through Norah.
Through most of the paper, each of these agents can observe all previous local decisions.\footnote{In Section~\ref{sec:incomplete-public}, we consider incomplete public signals.}  For example, Britta can observe Alexis's decision $\Hhat_1$; Carol can observe Britta's decision $\Hhat_2$ as well $\Hhat_1$; and so on.  We call the local decisions observed by other agents \emph{public signals}.

Later-acting agents have more degrees of freedom in choosing their decision thresholds because the decision thresholds may depend on the observed public signals.
Britta has two degrees of freedom:  she can apply different thresholds  when $\Hhat_1 = 0$ and when $\Hhat_1 = 1$.  Likewise, Carol has four degrees of freedom because she observes one of four public signal combinations $(\Hhat_1, \Hhat_2) \in \{(0, 0), (0, 1), (1, 0), (1, 1)\}.$  We use superscripts to specify the public signals that an agent observes.  For example, $\DecThres{2}{0}{}$ denotes Britta's decision threshold when she observes $\Hhat_1 = 0$ and $\DecThres{3}{01}{}$ denotes Carol's decision threshold when she observes $(\Hhat_1, \Hhat_2) = (0,1)$.
Furthermore, $\LocalISeq{2}{1}{}$ denotes Britta's Type I error probability when $\Hhat_1 = 1$ and $\LocalIISeq{3}{10}{}$ denotes Carol's Type II error probability when $(\Hhat_1, \Hhat_2) = (1,0)$.

In this paper, we will compare the optimal performance in the case when agents can observe public signals---called \emph{public voting}---to the case when they cannot---called \emph{secret voting}---so as to analyze to what extent the public signals are beneficial.

\section{Identical Agents}
\label{sec:IdenticalAgents}

Agents are assumed to observe conditionally iid private signals in this section.  We compare the optimal decision-making strategies in two cases:  public voting and secret voting.  By showing that the optimal decision thresholds are the same in these two cases, we will conclude that public signals are useless.

In secret voting, we can restrict the agents to use identical decision thresholds to simplify the problem~\cite{RhimG2013a}.  Using identical decision thresholds is asymptotically optimum for the binary hypothesis testing problem~\cite{Tsitsiklis1988}.  Furthermore, by numerical experiments, it turns out that constraining to identical decision rules causes little or no loss of performance for finite $N$ and the corresponding optimal fusion rule has the $L$-out-of-$N$ form~\cite{Tsitsiklis1993}.  Our numerical experiments
(not reported here)
for fixed $L$-out-of-$N$ fusion rules and conditionally iid private signals show that the optimal decision thresholds are in fact identical for any $N \leq 7$ for Gaussian and exponential likelihood functions.

\subsection{Complete Public Signals}

Suppose that agents can observe all decisions made by previous agents.  Specifically, Agent $n$ observes $\Hhat_1, \ldots, \Hhat_{n-1}$ before making her decision.

First, let us consider the simplest case of $N = 2$.
Alexis and Britta make a decision together and their fusion rule is the 1-out-of-2 (\textsc{or}) rule or the 2-out-of-2 (\textsc{and}) rule.
In secret voting, Alexis and Britta simultaneously make local decisions.  They use one decision threshold each.

On the other hand, in public voting Alexis first makes a decision and then Britta makes a decision upon observing Alexis's decision.
As always, Alexis has a single decision threshold.
Britta also has only one relevant decision threshold, like in secret voting.
While Britta seems to be able to choose two decision thresholds, one for $\Hhat_1 = 0$ and another for $\Hhat_1 = 1$, for each fusion rule there is one value of the public signal for which her decision becomes irrelevant to the team decision:
under the \textsc{or} rule, when $\Hhat_1 = 1$ the team decision is $\Hhat = 1$ regardless of Britta's local decision; and
under the \textsc{and} rule, when $\Hhat_1 = 0$ the team decision is $\Hhat = 0$ regardless of Britta's local decision.
Thus, both agents equally have one degree of freedom even when the voting is public.
This captures the basic idea behind public voting and secret voting yielding the same performance for $N=2$.
We provide a more technical proof below.
To draw attention to the distinction between the two cases, we use $\lambda$ to denote decision thresholds in secret voting and $\rho$ to denote those in public voting.

\vspace{0.1in}

\begin{theorem}
	The existence of public signals does not affect the optimal local decision rules for $N = 2$ under either of the two $L$-out-of-$N$ fusion rules.
	\label{thm:2Agent}
\end{theorem}

\vspace{0.1in}

\begin{IEEEproof}
	The fusion rule will be the 1-out-of-2 (\textsc{or}) rule or the 2-out-of-2 (\textsc{and}) rule.  Let us compare team Bayes risks in the secret voting and public voting cases under the \textsc{or} rule.
	
	In the secret voting scenario, the Bayes risk is given by
	\begin{equation}
		\Rs = c_{10} p_0 \left( \LocalI{1} + \left(1 - \LocalI{1}\right) \LocalI{2} \right) + c_{01} p_1 \LocalII{1} \LocalII{2},
		\label{eq:BR_1outof2_Secret}
	\end{equation}
	where Alexis's decision threshold $\DecThres{1}{}{}$ determines local error probabilities $\LocalI{1}$ and $\LocalII{1}$, and Britta's decision threshold $\DecThres{2}{}{}$ determines local error probabilities $\LocalI{2}$ and $\LocalII{2}$.  Their optimal decision thresholds $\lambda_1^\ast$ and $\lambda_2^\ast$ minimize \eqref{eq:BR_1outof2_Secret}.

    In the public voting scenario, the Bayes risk is in the same form
	\begin{equation}
		\Rp = c_{10} p_0 \left( \LocalISeq{1}{}{} + \left( 1 - \LocalISeq{1}{}{} \right) \LocalISeq{2}{0}{} \right) + c_{01} p_1 \LocalIISeq{1}{}{} \LocalIISeq{2}{0}{},
		\label{eq:BR_1outof2_Public}
	\end{equation}
	except that Britta's error probabilities, $\LocalISeq{2}{0}{}$ and $\LocalIISeq{2}{0}{}$, are controlled by her decision threshold $\rho_2^{_0}$ for $\Hhat_1 = 0$.
    Britta's decision threshold when $\Hhat_1 = 1$, $\rho_2^{_1}$, is irrelevant; thus, we can assume that $\rho_2^{_1} = \rho_2^{_0}$ without loss of optimality.
	
	Expressions~\eqref{eq:BR_1outof2_Secret} and~\eqref{eq:BR_1outof2_Public} are very similar;
	the only difference is the replacement of $(\LocalI{2},\LocalII{2})$ in~\eqref{eq:BR_1outof2_Secret}
	by $(\LocalISeq{2}{0}{},\LocalIISeq{2}{0}{})$ in~\eqref{eq:BR_1outof2_Public}.
	Now note that the set of achievable values for $(\LocalI{2},\LocalII{2})$ and $(\LocalISeq{2}{0}{},\LocalIISeq{2}{0}{})$ are identical;
	they are achieved by varying Britta's decision threshold ($\lambda_2$ or $\rho_2^{_0}$) in precisely the same local decision-making problem.
	Therefore, minimizing $\Rs$ and $\Rp$ results in equal Bayes risks, and these are achieved with decision thresholds satisfying the following:
	\begin{equation}
	    \rho_1^\ast = \lambda_1^\ast, \qquad \rho_2^{_0\ast} = \lambda_2^\ast.
    \end{equation}
	Since $\lambda_1^\ast = \lambda_2^\ast$, we also have $\rho_1^\ast = \rho_2^{_0\ast}$.
	Therefore, Alexis and Britta should not change their decision thresholds depending on whether or not the voting is public.
	
	The proof for the \textsc{and} fusion rule is similar.  While the precise expressions for $\Rs$ and $\Rp$ change, again we find that the achievable set of local Type~I and Type~II error probabilities are identical under secret and public voting, so the minima and optimum decision thresholds are unaffected by the public signal.
\end{IEEEproof}

\vspace{0.1in}

For larger $N$, we prove the lack of usefulness of the public signal by mathematical induction.

\vspace{0.1in}

\begin{theorem}
	Suppose that sharing local decisions does not change Alexis's decision threshold (i.e., $\rho_1^\ast = \lambda_1^\ast$).  If the optimal decision-making rules are the same in public voting and in secret voting for a specific $N$ and any $K$-out-of-$N$ fusion rule, then it is also true for a team of $N{+}1$ agents and any $L$-out-of-$(N{+}1)$ fusion rule.
	\label{thm:NtoN+1}
\end{theorem}

\vspace{0.1in}

\begin{IEEEproof}
    First, we consider the secret voting scenario with $N{+}1$ agents.  Since Agent $n$'s decision is critical only if the other $N$ local decisions are $L{-}1$ ones and $N{-}L{+}1$ zeros, the optimal decision threshold $\lambda_n^*$ is the solution to\footnote{Please see~\cite{RhimVG2012c} for a detailed description of how \eqref{eq:DecThres,N+1agents,parallel} is derived.}
    \begin{align}
        \frac{f_{Y_n | H} (\lambda_n \MID 1)}{f_{Y_n | H} (\lambda_n \MID 0)}
        & = \frac{c_{10} p_0 \binom{N}{L-1} \left(\LocalIx \right)^{L-1} \left(1 - \LocalIx \right)^{N-L+1}}{c_{01} p_1 \binom{N}{N-L+1} \left(\LocalIIx \right)^{N-L+1} \left(1 - \LocalIIx \right)^{L-1}} \nonumber \\
        & = \frac{c_{10} p_0 \left(\LocalIx \right)^{L-1} \left(1 - \LocalIx \right)^{N-L+1}}{c_{01} p_1 \left(\LocalIIx \right)^{N-L+1} \left(1 - \LocalIIx \right)^{L-1}},
        \label{eq:DecThres,N+1agents,parallel}
    \end{align}
    where we use that $\LocalI{1} = \LocalI{2} = \cdots = \LocalIx$ and $\LocalII{1} = \LocalII{2} = \cdots = \LocalIIx$ because the optimal decision thresholds of all agents are identical in secret voting.

    Next, in the public voting scenario, we can classify error cases depending on Alexis's detection result, e.g., when the true state is 0 and Alexis's decision is correct ($\Hhat_1 = 0$), a false alarm occurs if at least $L$ out of the remaining $N$ agents vote for 1.  The Bayes risk is given by

    {\small
    \begin{align}
        \Rp
        & = c_{10} p_0 \left( 1 - \LocalI{1} \right) \P\left\{\textstyle \sum_{n = 2}^{N+1} \Hhat_n \geq L \MID \Hhat_1 = H = 0\right\} \nonumber \\
        & \quad + c_{10} p_0 \LocalI{1} \P\left\{\textstyle \sum_{n = 2}^{N+1} \Hhat_n \geq L - 1 \MID \Hhat_1 = 1, H = 0\right\} \nonumber \\
        & \quad + c_{01} p_1 \LocalII{1} \P\left\{\textstyle \sum_{n = 2}^{N+1} \Hhat_n \leq L - 1 \MID \Hhat_1 = 0, H = 1\right\} \nonumber \\
        & \quad + c_{01} p_1 \left( 1 - \LocalII{1} \right) \P\left\{ \sum_{n = 2}^{N+1} \Hhat_n \leq L - 2 \MID \Hhat_1 = H = 1\right\} \nonumber \\
        & \triangleq R_0 \left(p_0 (1 - \LocalISeq{1}{}{} ) + p_1 \LocalIISeq{1}{}{} \right) + R_1 \left(p_0 \LocalISeq{1}{}{} + p_1 ( 1 -  \LocalIISeq{1}{}{} ) \right),
        \label{eq:BR,L-out-of-N+1,serial}
    \end{align}
    }where $R_0$ and $R_1$ are specified in \eqref{eq:BR-B0} and \eqref{eq:BR-B1} and we define
    \begin{align}
        \Belief^{_0} & \triangleq \frac{p_0 \left(1 - \LocalISeq{1}{}{} \right)}{p_0 \left(1 - \LocalISeq{1}{}{} \right) + p_1 \LocalIISeq{1}{}{}} = \P\{H = 0 \MID \Hhat_1 = 0\}, \nonumber \\
        \Belief^{_1} & \triangleq \frac{p_0 \LocalISeq{1}{}{}}{p_0 \LocalISeq{1}{}{} + p_1 \left( 1 -  \LocalIISeq{1}{}{} \right)} = \P\{H = 0 \MID \Hhat_1 = 1\}.
        \label{eq:BeliefUpdate_N+1}
    \end{align}

    When Agents $2, 3, \ldots, N{+}1$ observe $\Hhat_1 = 0$, their optimal decision strategy is to minimize the term $R_0$ from \eqref{eq:BR,L-out-of-N+1,serial}:
    \begin{align}
        R_0 & = c_{10} \Belief^{_0} \P\left\{\textstyle \sum_{n = 2}^{N+1} \Hhat_n^{_0} \geq L \MID H = 0\right\}  \nonumber \\
        & \quad + c_{01}(1 - \Belief^{_0}) \P\left\{\textstyle \sum_{n = 2}^{N+1} \Hhat_n^{_0} \leq L - 1 \MID H = 1\right\},
        \label{eq:BR-B0}
    \end{align}
    where the condition $\Hhat_1 = 0$ is embedded in the superscript of $\Hhat_n^{_0}$.  Please note that $R_0$ is the same as the Bayes risk of $N$ agents when the prior probability is $\Belief^{_0}$ and the fusion is done by the $L$-out-of-$N$ rule.  It implies that the optimal decision thresholds of Agents $2, 3, \ldots, N{+}1$ are the same as those of $N$ agents with prior probability $\Belief^{_0}$ and the $L$-out-of-$N$ fusion rule.

    Likewise, when Agents $2, 3, \ldots, N{+}1$ observe $\Hhat_1 = 1$, their optimal decision strategy is to minimize the term $R_1$ from \eqref{eq:BR,L-out-of-N+1,serial}:
    \begin{align}
        R_1 & = c_{10} \Belief^{_1} \P\left\{\textstyle \sum_{n = 2}^{N+1} \Hhat_n^{_1} \geq L - 1 \MID H = 0\right\}  \nonumber \\
        & \quad + c_{01}(1 - \Belief^{_1}) \P\left\{\textstyle \sum_{n = 2}^{N+1} \Hhat_n^{_1} \leq L - 2 \MID H = 1\right\}.
        \label{eq:BR-B1}
    \end{align}
    Their optimal decision thresholds are the same as those of $N$ agents with prior probability $\Belief^{_1}$ and the $(L{-}1)$-out-of-$N$ fusion rule.  Fig.~\ref{fig:EvolutionN} depicts the evolution of the problem corresponding to Alexis's decision $\Hhat_1$.

    \begin{figure}
        \centering
        \includegraphics[width=3.4in]{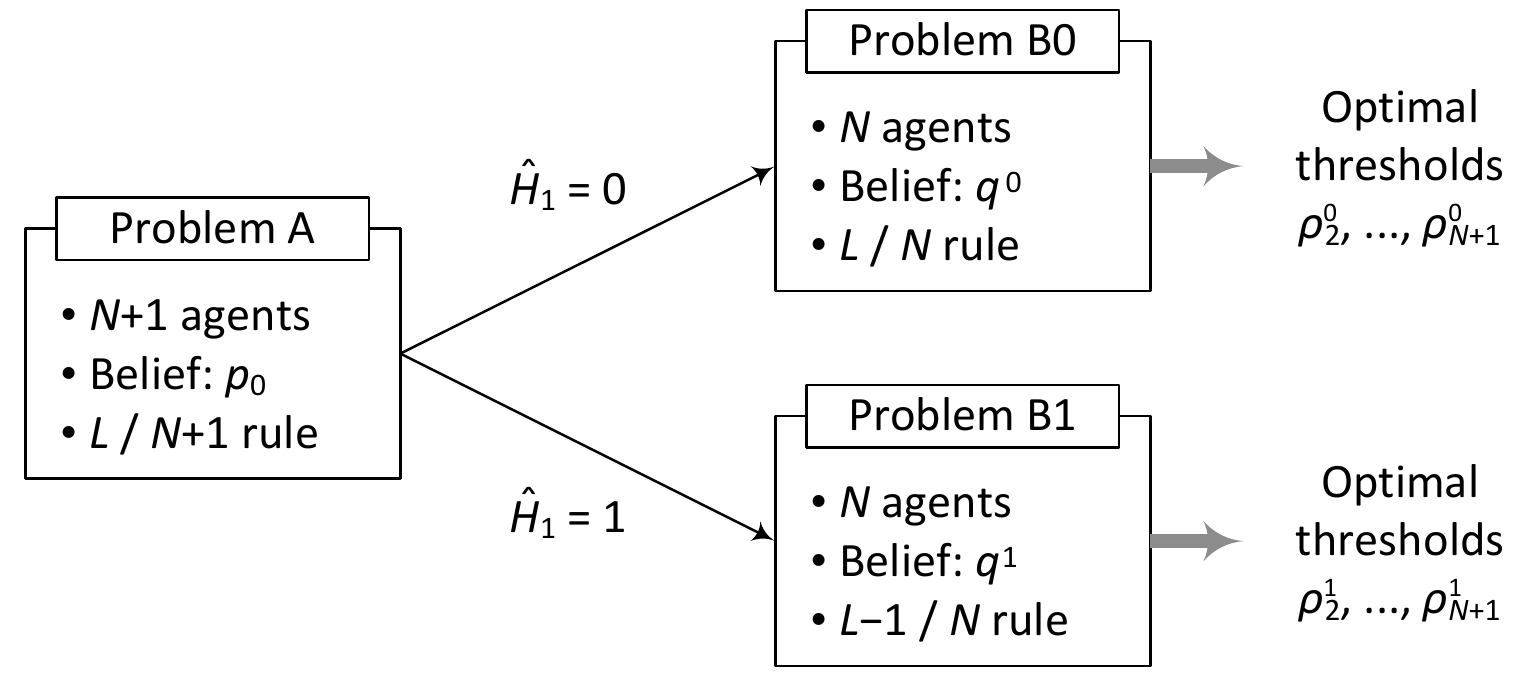}
        \caption{An $(N{+}1)$-agent problem is divided into two $N$-agent problems depending on Alexis's decision $\Hhat_1$.}
        \label{fig:EvolutionN}
    \end{figure}

    Let us find the optimal thresholds $\rho_2^{_0 \ast}, \rho_3^{_0 \ast}, \ldots, \rho_{N+1}^{_0 \ast}$ in Problem~B0 in Fig.~\ref{fig:EvolutionN}.  In fact, Problem~B0 is also a public voting scenario; agents observe $\Hhat_2$, $\Hhat_3$, and so on.  However, because of the assumption that the existence of the public signals does not affect optimal decision thresholds of a team of $N$ agents for any $K$-out-of-$N$ fusion rule, we can find the optimal thresholds as if the agents do secret voting.  Since an agent's decision is critical only if the other $N{-}1$ local decisions consist of $L{-}1$ ones and $N{-}L$ zeros, the optimal decision threshold $\rho^{_0 \ast}$ is the solution to

    \vspace{-3mm}
    {\small
    \begin{align}
        \frac{f_{Y | H} (\rho^{_0} \MID 1)}{f_{Y | H} (\rho^{_0} \MID 0)}
        & = \frac{c_{10} \Belief^{_0} \binom{N-1}{L-1} \left(\LocalIxSeq{0} \right)^{L-1} \left(1{-}\LocalIxSeq{0} \right)^{N-L}}{c_{01} (1{-}\Belief^{_0}) \binom{N-1}{N-L} \left(\LocalIIxSeq{0} \right)^{N-L} \left(1{-}\LocalIIxSeq{0} \right)^{L-1}} \nonumber \\
        & = \frac{c_{10} p_0 \left(1{-}\LocalIx \right) \left(\LocalIxSeq{0} \right)^{L-1} \left(1{-}\LocalIxSeq{0} \right)^{N-L}}{c_{01} p_1 \LocalIIx \left(\LocalIIxSeq{0} \right)^{N-L} \left(1{-}\LocalIIxSeq{0} \right)^{L-1}},
        \label{eq:DecThres0,Nagents,serial}
    \end{align}}where $\Belief^{_0}$ is replaced by \eqref{eq:BeliefUpdate_N+1}.  Due to the assumption that $\rho_1^{\ast} = \lambda_1^{\ast}$, $\LocalIx$ and $\LocalIIx$ in \eqref{eq:DecThres0,Nagents,serial} are the same as $\LocalIx$ and $\LocalIIx$ in \eqref{eq:DecThres,N+1agents,parallel}.

    Comparing \eqref{eq:DecThres0,Nagents,serial} to \eqref{eq:DecThres,N+1agents,parallel}, we can find that they have the same solutions, i.e., $\rho_i^{_0 \ast} = \lambda_i^{\ast}$.  Therefore, the agents should not change their decision thresholds after observing $\Hhat_1 = 0$.

    We can also find the optimal thresholds $\rho_2^{_1 \ast}, \ldots, \rho_{N+1}^{_1 \ast}$ in Problem~B1 in Fig.~\ref{fig:EvolutionN} by looking at the $N$-agent problem without public signals:

    \vspace{-3mm}
    {\small
    \begin{align}
        \frac{f_{Y | H} (\rho^{_1} \MID 1)}{f_{Y | H} (\rho^{_1} \MID 0)}
        & = \frac{c_{10} \Belief^{_1} \binom{N-1}{L-2} \left(\LocalIxSeq{1} \right)^{L-2} \left(1{-}\LocalIxSeq{1} \right)^{N-L+1}}{c_{01} (1{-}\Belief^{_1}) \binom{N-1}{N-L+1} \left(\LocalIIxSeq{1} \right)^{N-L+1} \left(1{-}\LocalIIxSeq{1} \right)^{L-2}} \nonumber \\
        & = \frac{c_{10} p_0 \LocalIx \left(\LocalIxSeq{1} \right)^{L-2} \left(1{-}\LocalIxSeq{1} \right)^{N-L+1}}{c_{01} p_1 \left(1{-}\LocalIIx\right) \left(\LocalIIxSeq{1} \right)^{N{-}L{+}1} \left(1{-}\LocalIIxSeq{1} \right)^{L{-}2}}.
        \label{eq:DecThres1,Nagents,serial}
    \end{align}}Again, due to the assumption that $\rho_1^{\ast} = \lambda_1^{\ast}$, $\LocalIx$ and $\LocalIIx$ in \eqref{eq:DecThres1,Nagents,serial} are the same as $\LocalIx$ and $\LocalIIx$ in \eqref{eq:DecThres,N+1agents,parallel}.  We reach the same conclusion that the two equations have the same solutions, i.e., $\rho_i^{_1 \ast} = \lambda_i^{\ast}$, by comparing \eqref{eq:DecThres1,Nagents,serial} to \eqref{eq:DecThres,N+1agents,parallel}.  Thus, the agents should not change their decision thresholds after observing $\Hhat_1 = 1$.

    Consequently, for a team of $N{+}1$ agents and any $L$-out-of-$(N{+}1)$ rule, their optimal decision thresholds are the same whether they observe previous decisions or not.
\end{IEEEproof}

\vspace{0.1in}

\begin{corollary}
    Suppose that sharing local decisions does not change Alexis's decision rule (i.e., $\rho_1^{\ast} = \lambda_1^{\ast}$).  For any $N$ and $L$-out-of-$N$ fusion rule, the existence of the public signals does not affect optimal decision thresholds of a team of $N$ agents.
    \label{cor:GeneralN}
\end{corollary}

\vspace{0.1in}

\begin{IEEEproof}
    Use mathematical induction with Theorems~\ref{thm:2Agent} and~\ref{thm:NtoN+1}.
\end{IEEEproof}

\vspace{0.1in}

This result requires the assumption that Alexis uses the same decision rule in both secret and public voting.  This assumption is trivially true for $N = 1$ and also true for $N = 2$ by proof.  In addition, our numerical experiments for $N \leq 9$ confirmed that it is true.  Thus, this assumption seems heuristically true.  In particular, we fail to see how Alexis would choose between increasing or decreasing her decision threshold based on the existence of public signals.

\emph{Individual correctness vs.\ team correctness, and belief update vs.\ fusion rule evolution:}
One may have the intuition that, because the public signals are not independent of $H$, it cannot make sense for agents to ignore the public signals.  Indeed, if the goal of Agent $n$ is to make a correct decision herself (ignoring her role in the team) and she observes the public signal of Agent $i$ before making her decision, she should not ignore $\Hhat_i$.
If $\Hhat_i = 0$, then Agent $n$ would include the information in her LRT to make a better choice as follows:
\begin{equation}
	\frac{P_{Y_n, \Hhat_i \MID H}(y_n, 0 \MID 1)}{P_{Y_n, \Hhat_i \MID H}(y_n, 0 \MID 0)}\overset{\Hhat_n = 1}{\underset{\Hhat_n = 0}{\gtreqless}} \frac{c_{10} p_0}{c_{01} p_1}.
	\label{eq:ModifiedLRT}
\end{equation}
Since Agent $n$'s private signal $Y_n$ and Agent $i$'s private signal $Y_i$ are independent given $H$, so are $Y_n$ and $\Hhat_i$.  Then~\eqref{eq:ModifiedLRT} is transformed as
\begin{align}
	\frac{f_{Y_n \MID H}(y_n \MID 1)}{f_{Y_n \MID H}(y_n \MID 0)}
	& \overset{\Hhat_n = 1}{\underset{\Hhat_n = 0}{\gtreqless}} \frac{c_{10} p_0 \P\{\Hhat_i = 0 \MID H = 0\}}{c_{01} p_1 \P\{\Hhat_i = 0 \MID H = 1\}} \nonumber \\
	& \ \ = \ \ \frac{c_{10} p_0 \P\{H = 0 \MID \Hhat_i = 0\}}{c_{01} p_1 \P\{H = 1 \MID \Hhat_i = 0\}} = \frac{c_{10} \Belief^{_0}}{c_{10} (1 - \Belief^{_0})},
\end{align}
where $\Belief^{_0} = \P\{H = 0 \MID \Hhat_i = 0\}$.
Agent $n$'s LRT with two observations $Y_n$ and $\Hhat_i$ becomes an LRT with one observation $Y_n$ as if she updates the prior probability of $H = 0$ from $p_0$ to $\Belief^{_0}$ upon observing $\Hhat_i = 0$.
More generally, the updated belief of $H = h$ has a form of $\P\{H = h \MID \mbox{observed public signals}\}$, assuming the agents observe conditionally independent private signals.  The updated belief of $H = 0$ will be higher than the prior probability if previous agents have chosen 0 and lower if they have chosen $1$.

Return now to the optimization of team performance.
In the proof of Theorem~\ref{thm:NtoN+1}, \eqref{eq:BR-B0} and \eqref{eq:BR-B1} reveal that the belief is updated in the same manner when the agents perform group decision making.
The difference is that these formulas contain \emph{fusion rule evolution} (depicted in Fig.~\ref{fig:EvolutionN}) along with the \emph{belief update}.
While the belief update is common in the social learning literature---in fact, central to it---the evolution of the fusion rule hardly appears in the literature.

\begin{figure}
    \centering
    \includegraphics[width=3.4in]{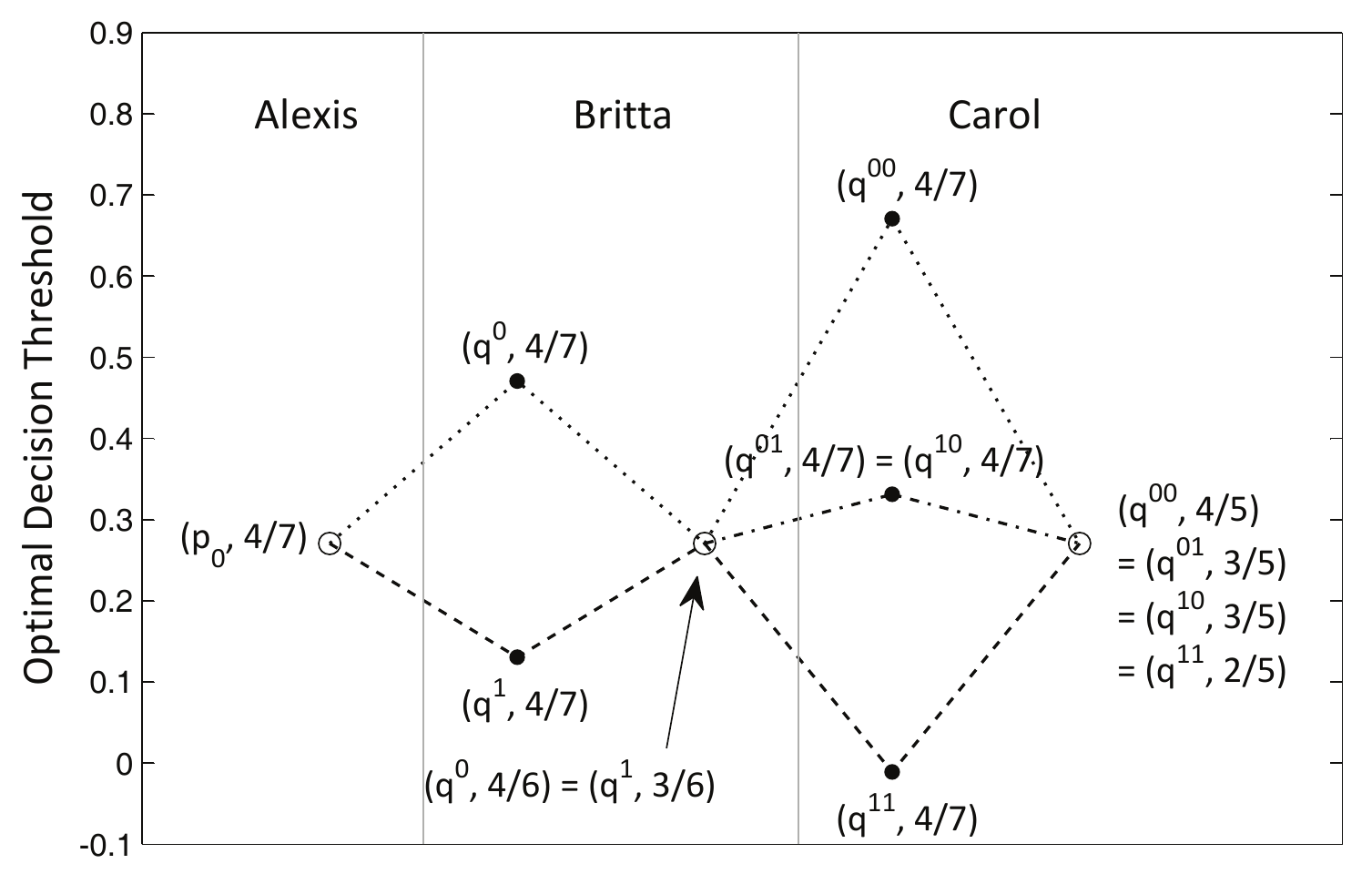}
    \caption{Illustration of the separate effects of belief updates and fusion rule updates.
    The diagram depicts decision thresholds for the first three agents in an example with $p_0 = 0.25$ and the 4-out-of-7 fusion rule.
    Agents observe $H \in \{0,1\}$ corrupted by iid Gaussian additive noises with zero mean and unit variance, and $c_{10} = c_{01} = 1$.
    The $(p,L/N)$ label on a marker ($\bullet$ or $\circ$) indicates that its height represents the optimal decision threshold for prior probability $p$ and the $L$-out-of-$N$ fusion rule.
    Alexis has an initial decision threshold depending only on the prior probability $p_0$ and the 4-out-of-7 fusion rule (leftmost $\circ$).
    If Britta considers belief updates only, the optimal decision threshold is changed from Alexis's decision threshold to a new value that depends on $\Hhat_1$ (two leftmost $\bullet$'s).
    However, after adopting the fusion rule evolution as well, Britta's optimal decision threshold returns to equal Alexis's decision threshold (center $\circ$).
    Similarly, if Carol considers belief updates only, the four values for $(\Hhat_1,\Hhat_2)$ lead to three distinct decision thresholds (three rightmost $\bullet$'s).
    After accounting for the fusion rule evolution, Carol's optimal decision threshold returns to equal Alexis's decision threshold (rightmost $\circ$).
}
    \label{fig:DecThresChange}
\end{figure}

The fusion rule evolution arises because, in a public voting scenario, an agent should hesitate to invalidate the votes of later-acting agents teammate should  of aggregation by voting and the agents' observations of all previous decisions.
Each agent can keep a running tally of the numbers of 0 and 1 votes.
A large number of 1 votes implies that only a few more 1 votes will determine the team decision to be 1 even before all agents vote;
thus, in order to vote 1, an agent should need a strong private signal in support of 1 to take the risk of making later-acting agents' votes null.
By Corollary~\ref{cor:GeneralN}, the effects of the belief update and fusion rule evolution cancel exactly.
A numerical example is detailed in Fig.~\ref{fig:DecThresChange}.
Reading that figure from left to right, for each agent after Alexis, the belief update is done first and then the fusion rule evolution brings the optimal decision threshold back exactly to the optimal decision threshold of Alexis.

In conclusion, while belief updates encourage later-acting agents to agree with the earlier acting agents (the phenomenon that classically causes herding), decision-making by voting causes fusion rule evolution as well.
This discourages agreement of later-acting agents with the earlier-acting agents.
By proving that the optimal thresholds of the last $N$ agents are the same in the cases when $\Hhat_1 = 0$, when $\Hhat_1 = 1$, and when they do not know $\Hhat_1$, it is proved that the effects of the former and the latter are exactly canceled out.

\subsection{Incomplete Public Signals}
\label{sec:incomplete-public}

\begin{figure}
    \centering{
    \subfloat[]{\includegraphics[width=3.3in]{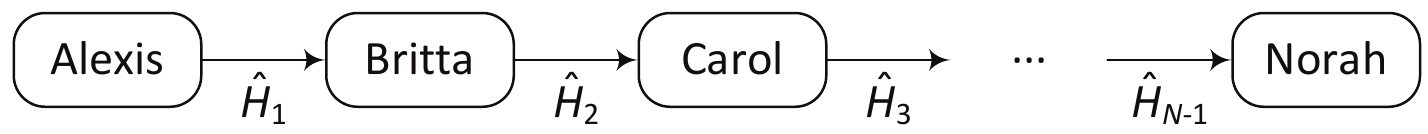} \label{fig:PartialPublicVoting1}}
    \hfill
    \subfloat[]{\includegraphics[width=2.1in]{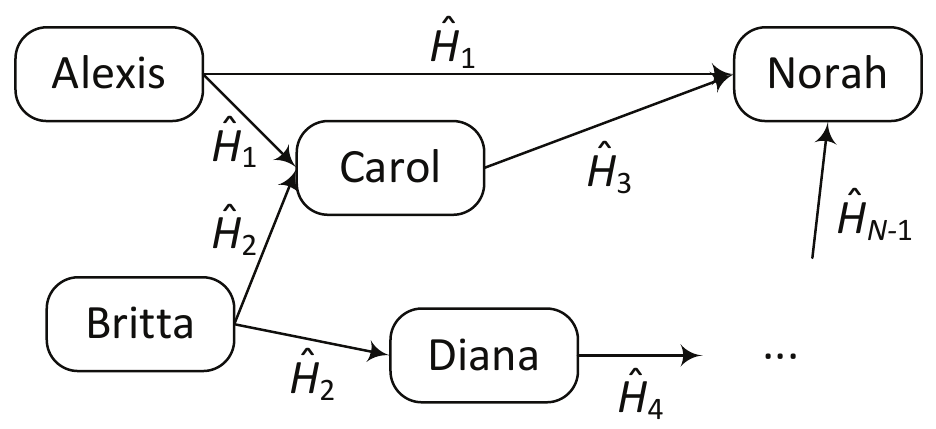} \label{fig:PartialPublicVoting2}}
    }
    \caption{Examples of partial public voting.  (a) Each agent sequentially observes the decision made by the agent right before her.  (b) Each agent observes the decisions made by her neighbors.}
    \label{fig:PartialPublicVoting}
\end{figure}

We now extend our analysis to scenarios in which each agent is aware of an arbitrary subset of the previously made decisions.
For example, each agent may observe the public signal only from its neighbors in a sequence (see Fig.~\ref{fig:PartialPublicVoting1})
or the communication topology may be more arbitrary (see Fig.~\ref{fig:PartialPublicVoting2}).
Let us say that agents perform \emph{partial public voting} when agents observe proper subsets of precedent local decisions.

\vspace{0.1in}

\begin{corollary}
	Observing a subset of public signals does not affect optimal decision rules and performance of a team.
\end{corollary}

\vspace{0.1in}

\begin{IEEEproof}
	First, we will show that a team of agents observing only a subset of public signals (Team~A) cannot outperform a team of the same size that consists of those who observe full public signals (Team~B).  The proof is by contradiction.  Let us assume that Team~A can outperform Team~B using their optimal decision strategy.  Then, since what each agent in Team~A observes is also observed by the corresponding agents in Team~B, Team~B can mimic the optimal strategy of Team~A\@.  For Agent $n$ in Team~B, all she has to do is ignore the public signals that Agent $n$ in Team~A cannot observe.  After mimicking the strategy of Team~A, the performance of Team~B becomes the same as that of Team~A\@.  This contradicts our assumption.  Hence Team~A cannot outperform Team~B\@.
	
	Next, let us consider Team~A and a team of agents not observing any public signals (Team~C).  We can prove that Team~C cannot outperform Team~A through similar logic.
	
	Corollary~\ref{cor:GeneralN} implies that Team~C in fact performs as well as Team~B\@.  Therefore Team~A is also as good as Teams~B and~C with respect to their optimal performances.
\end{IEEEproof}

\vspace{0.1in}

The convenience of secret voting emerges especially when agents cannot observe all public signals.  Even though partial public voting cannot outperform public voting, the former requires more computations for Bayesian learning.  When the agents observe a subset of public signals, they need to consider all possible realizations of the public signals that they cannot observe in order to perform Bayesian learning.  For example, in Fig.~\ref{fig:PartialPublicVoting1}, Carol observes Britta's decision but not Alexis's decision.   Her updated belief when $\Hhat_2 = 0$ will be computed as follows:
\begin{align}
	\Belief_3^{_0} & = \P\{H = 0 \MID \Hhat_2 = 0\} \nonumber \\
	& = \sum_{\hhat_1} \P\{H = 0 \MID \Hhat_2 = 0, \Hhat_1 = \hhat_1\} \P\{\Hhat_1 = \hhat_1 \MID \Hhat_2 = 0\} \nonumber \\
	& = \sum_{\hhat_1} \frac{ \P\{H = 0 \MID \Hhat_2 = 0, \Hhat_1 = \hhat_1\} \P\{\Hhat_1 = \hhat_1, \Hhat_2 = 0\}}{\P\{\Hhat_1 = 0, \Hhat_2 = 0\} + \P\{\Hhat_1 = 1, \Hhat_2 = 0\}} .
\end{align}
This process is more complicated than belief update with the knowledge of both $\Hhat_1 = 0$ and $\Hhat_2 = 0$, which is just to compute $\P\{H = 0 \MID \Hhat_2 = 0, \Hhat_1 = 0\}$.

Instead of accepting this complexity, the agents should ignore the public signals.
Since the optimal secret voting strategy performs equally to the optimal public voting strategy, it is economical for them to not share any public signals at all.

\section{Agents with Different Likelihoods}
\label{sec:Experts-Crowd}

Agents may have private signals that relate differently to the hypothesis.  As one example, this occurs when we model each agent's private signal as the hypothesis observed over an additive noise channel, and the
channels have different SNRs.
We could think of the agents with relatively high SNRs as \emph{experts} and those with low SNRs as \emph{novices}.
Their decisions are not equally informative, unlike in the identical-agent case of Section~\ref{sec:IdenticalAgents}.  We will show that the public signals are futile in cases where the fusion rule requires unanimity but useful in other cases.

Now that the agents' private signals do not have the same distributions, the order in which the agents act matters to their team performance.  To distinguish the agents, we name them in descending order of SNRs:  Amy has the highest SNR, Beth has the second-highest SNR, and so on.  However, we keep the notation that the numbering of agents (and hence the subscript indices) indicate the order of decision making, i.e., Agent 1 acts first, but may or may not be Amy.

\subsection{\textsc{And} and \textsc{or} Fusion Rules}
\label{sec:Experts-Crowd_AND/OR}

The \textsc{and} ($N$-out-of-$N$) and \textsc{or} ($1$-out-of-$N$) rules have a common feature.  The team decision requires unanimity of the agents of one type or the other: for a team decision of 1 under the \textsc{and} rule, the agents must unanimously decide 1; and for a team decision of 0 under the \textsc{or} rule, they must unanimously decide 0\@.
We thus call these \emph{unanimity fusion rules}.
This characteristic gives a special result for these fusion rules.

Let us consider a team of two agents with the \textsc{or} rule: Amy with a higher SNR and Beth with a lower SNR\@.  From the discussion of the two-agent case in Section~\ref{sec:IdenticalAgents}, both Amy and Beth have one degree of freedom.  Suppose Amy makes her decision first with decision threshold $\DecThres{\rm A}{}{}$, and Beth then makes her decision with decision threshold $\DecThres{\rm B}{0}{}$, regardless of Amy's decision.  Their minimum Bayes risk is
\begin{equation}
    \min_{\DecThres{\rm A}{}{}, \DecThres{\rm B}{0}{}} c_{10} p_0 \left(\LocalISeq{\rm A}{}{} + \left(1 - \LocalISeq{\rm A}{}{}\right) \LocalISeq{\rm B}{0}{}\right) + c_{01} p_1 \LocalIISeq{\rm A}{}{} \LocalIISeq{\rm B}{0}{}.
    \label{eq:MinBR_OR_AB}
\end{equation}

Now suppose that they switch their positions: Beth first makes her decision with decision threshold $\rho_{\rm B}^{_{\ }}$, and Amy then makes her decision with decision threshold $\rho_{\rm A}^{_0}$, regardless of Beth's decision.  Their minimum Bayes risk is now given by
\begin{equation}
    \min_{\rho_{\rm B}^{_{\ }}, \rho_{\rm A}^{_0}} c_{10} p_0 \left(\LocalISeq{\rm B}{}{} + \left(1 - \LocalISeq{\rm B}{}{}\right) \LocalISeq{\rm A}{0}{}\right) + c_{01} p_1 \LocalIISeq{\rm B}{}{} \LocalIISeq{\rm A}{0}{}.
    \label{eq:MinBR_OR_BA}
\end{equation}

The situation is very similar to the proof of Theorem~\ref{thm:2Agent}.
We have two nearly identical expressions,
\eqref{eq:MinBR_OR_AB} and \eqref{eq:MinBR_OR_BA},
and each agent has a single decision threshold to vary.
This leads to identical minima achieved with identical decision thresholds.
Specifically, \eqref{eq:MinBR_OR_BA} can be rearranged to
\begin{equation}
    \min_{\rho_{\rm B}^{_{\ }}, \rho_{\rm A}^{_0}} c_{10} p_0 \left( \LocalISeq{\rm A}{0}{} + \left(1 - \LocalISeq{\rm A}{0}{}\right) \LocalISeq{\rm B}{}{} \right) + c_{01} p_1 \LocalIISeq{\rm A}{0}{} \LocalIISeq{\rm B}{}{}.
    \label{eq:MinBR_OR_BA_rearranged}
\end{equation}
Now the only difference between \eqref{eq:MinBR_OR_AB} and \eqref{eq:MinBR_OR_BA_rearranged} is the replacement of
$(\LocalISeq{\rm A}{}{}, \LocalIISeq{\rm A}{}{}, \LocalISeq{\rm B}{0}{}, \LocalIISeq{\rm B}{0}{})$ in \eqref{eq:MinBR_OR_AB} by
$(\LocalISeq{\rm A}{0}{}, \LocalIISeq{\rm A}{0}{}, \LocalISeq{\rm B}{}{}, \LocalIISeq{\rm B}{}{})$ in \eqref{eq:MinBR_OR_BA_rearranged}.
Each of these 4-tuples is the combinations of Type~I and Type~II error probabilities for Amy and Beth as each agent varies a single threshold;
thus, the achievable 4-tuples are identical.
Therefore, the optimal decision thresholds are also the same:
\begin{equation}
    \lambda_{\rm A}^\ast = \rho_{\rm A}^{_0\ast} \quad \mbox{and} \quad \lambda_{\rm B}^{_0\ast} = \rho_{\rm B}^\ast.
\end{equation}

In conclusion, not only is the public signal useless but also the agents' decision-making order does not affect the optimal team decision.  Their optimal strategy is just to adopt the decision thresholds $\lambda_1^*$ and $\lambda_2^*$ that minimize the Bayes risk:
\begin{align}
    (\lambda_1^*, \lambda_2^*) = \argmin_{(\DecThres{1}{}{}, \DecThres{2}{}{})} & \left\{ c_{10} p_0 \left(1 - \left(1 - \LocalI{1}\right) \left(1 - \LocalI{2}\right) \right) \right. \nonumber \\
    & \left. + c_{01} p_1 \LocalII{1} \LocalII{2} \right\}.
\end{align}

We can extend this result to $N$ agents as long as the fusion is performed by the \textsc{or} rule or the \textsc{and} rule.

\vspace{0.1in}

\begin{theorem}
    For a unanimity fusion rule, secret voting is the optimal strategy even when agents observe private signals with different SNRs.
    Specifically, public signals are useless and the ordering of agents does not affect their optimal decision rules nor the resulting performance.
\end{theorem}

\vspace{0.1in}

\begin{IEEEproof}
    For the \textsc{or} rule, where the Bayes risk is given by
    \begin{align}
        c_{10} p_0 \left( 1 - \prod_{n = 1}^{N} \left(1 - \LocalI{n}\right) \right)
        + c_{01} p_1 \prod_{n = 1}^{N} \LocalII{n}, \label{eq:ORFusionRule-IV}
    \end{align}
    each agent has a meaningful decision threshold only if all previous agents declare 0\@.  Otherwise, decisions of the remaining agents are irrelevant.  Thus, without loss of optimality, we can constrain that the agents optimize their decision thresholds $\DecThres{n}{}{}$ for the case when $\Hhat_1 = \Hhat_2 = \cdots = \Hhat_{n-1} = 0$ and use the same decision threshold for any public signals.  In fact, they need not know the public signals; they just need to perform decision making as if all public signals are 0.

    Furthermore, the symmetry in \eqref{eq:ORFusionRule-IV} implies that indices of Agents $m$ and $n$ are interchangeable.  Therefore, the ordering of agents does not affect the optimal values of decision thresholds and, consequently, the minimum Bayes risk.

    For the \textsc{and} rule, the Bayes risk is given by
    \begin{align}
        c_{10} p_0 \prod_{n = 1}^{N} \LocalI{n}
        + c_{01} p_1 \left( 1 - \prod_{n = 1}^{N} \left(1 - \LocalII{n}\right) \right),
        \label{eq:ANDFusionRule-IV}
    \end{align}
    and we can prove the statement in a similar way.
\end{IEEEproof}

\vspace{0.1in}

In general, the solution to minimizing~\eqref{eq:ORFusionRule-IV} or~\eqref{eq:ANDFusionRule-IV} does not satisfy $\lambda_1^\ast = \lambda_2^\ast = \cdots = \lambda_N^\ast$, unlike in Section~\ref{sec:IdenticalAgents}.  Agents will use different optimal decision thresholds corresponding to their SNRs.  We are arguing not that the agents should have the same decision thresholds but that they should keep their decision thresholds the same in any decision-making position.  Furthermore, each agent should make a decision sincerely based on her private signal and should not care about public signals because her vote will count only when all the other agents vote in a particular way (0 for the \textsc{or} rule or 1 for the \textsc{and} rule).

\subsection{Other Fusion Rules}

Optimal decision making is more complex for other fusion rules due to the increase of degrees of freedom.  Even in the simplest case when three agents make a decision with the \textsc{majority} (2-out-of-3) rule, the last two agents have two meaningful degrees of freedom each.  The second agent can have different decision thresholds for $\Hhat_1 = 0$ ($\DecThres{2}{0}{}$) and for $\Hhat_1 = 1$ ($\DecThres{2}{1}{}$), and the last agent can also have different ones for $(\Hhat_1,\Hhat_2) = (0,1)$ ($\DecThres{3}{01}{}$) and for $(\Hhat_1,\Hhat_2) = (1,0)$ ($\DecThres{3}{10}{}$).
(The third agent is irrelevant for $(\Hhat_1,\Hhat_2) = (0,0)$ or $(\Hhat_1,\Hhat_2) = (1,1)$ because the team decision has been made without her decision.)
Learning from public signals can be helpful in decision making due to these extra degrees of freedom, unlike in Section~\ref{sec:Experts-Crowd_AND/OR}.

Our symmetric fusion rule always treats all local decisions with equal weights even though they are made by agents that experience different SNRs.  Thus, the team decision can be improved by social learning, which inevitably unbalances weights of local decisions.

\begin{figure}
    \centering
    \includegraphics[width=3.4in]{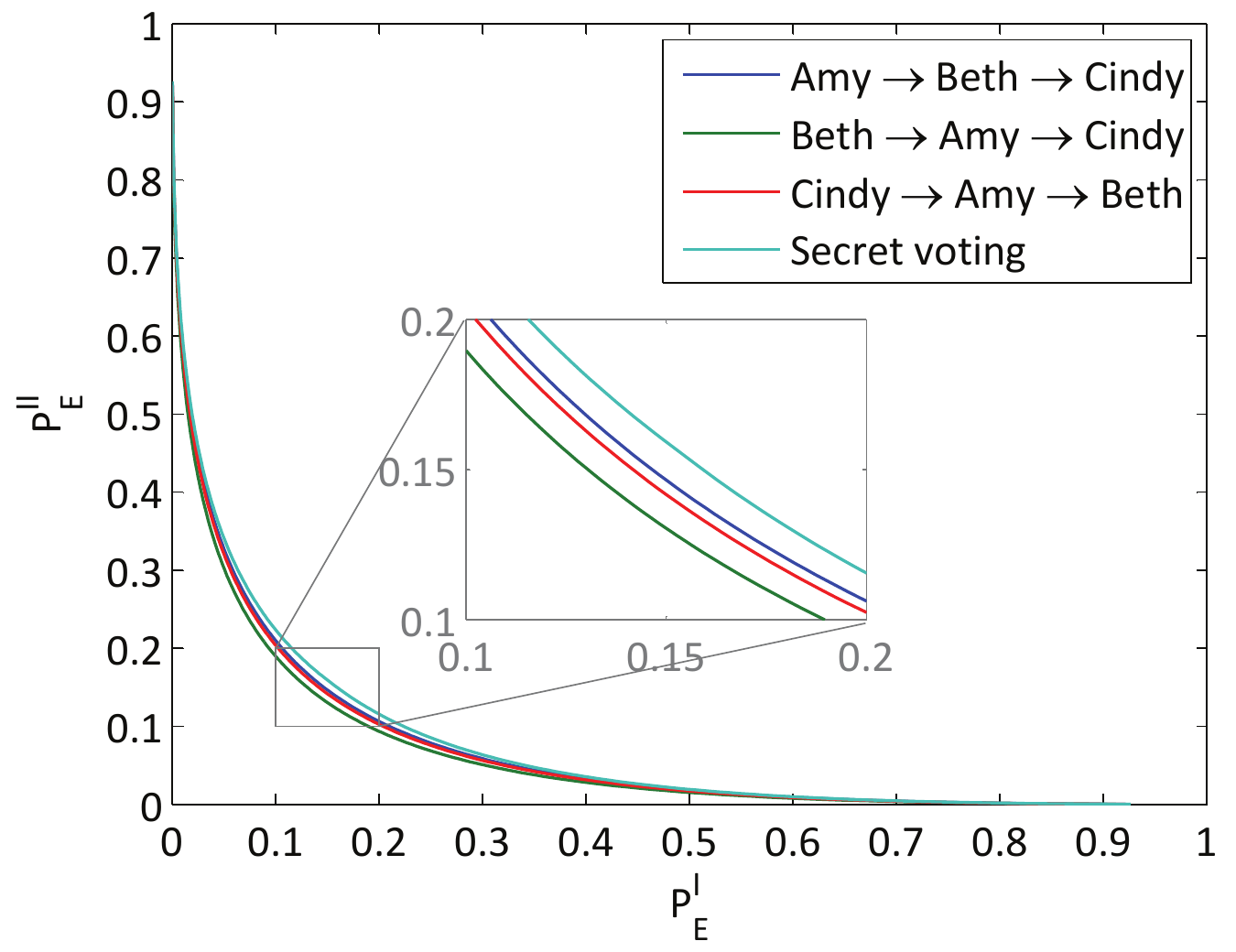}
    \caption{Lower bounds of operating regions for different orderings of three agents, Amy ($\sigma_{\rm A}^2 = 0.25$), Beth ($\sigma_{\rm B}^2 = 1$), and Cindy ($\sigma_{\rm C}^2 = 2.25$), and the \textsc{majority} fusion rule.}
    \label{fig:ROC_2outof3}
\end{figure}

\begin{figure}
    \centering
    \includegraphics[width=3.4in]{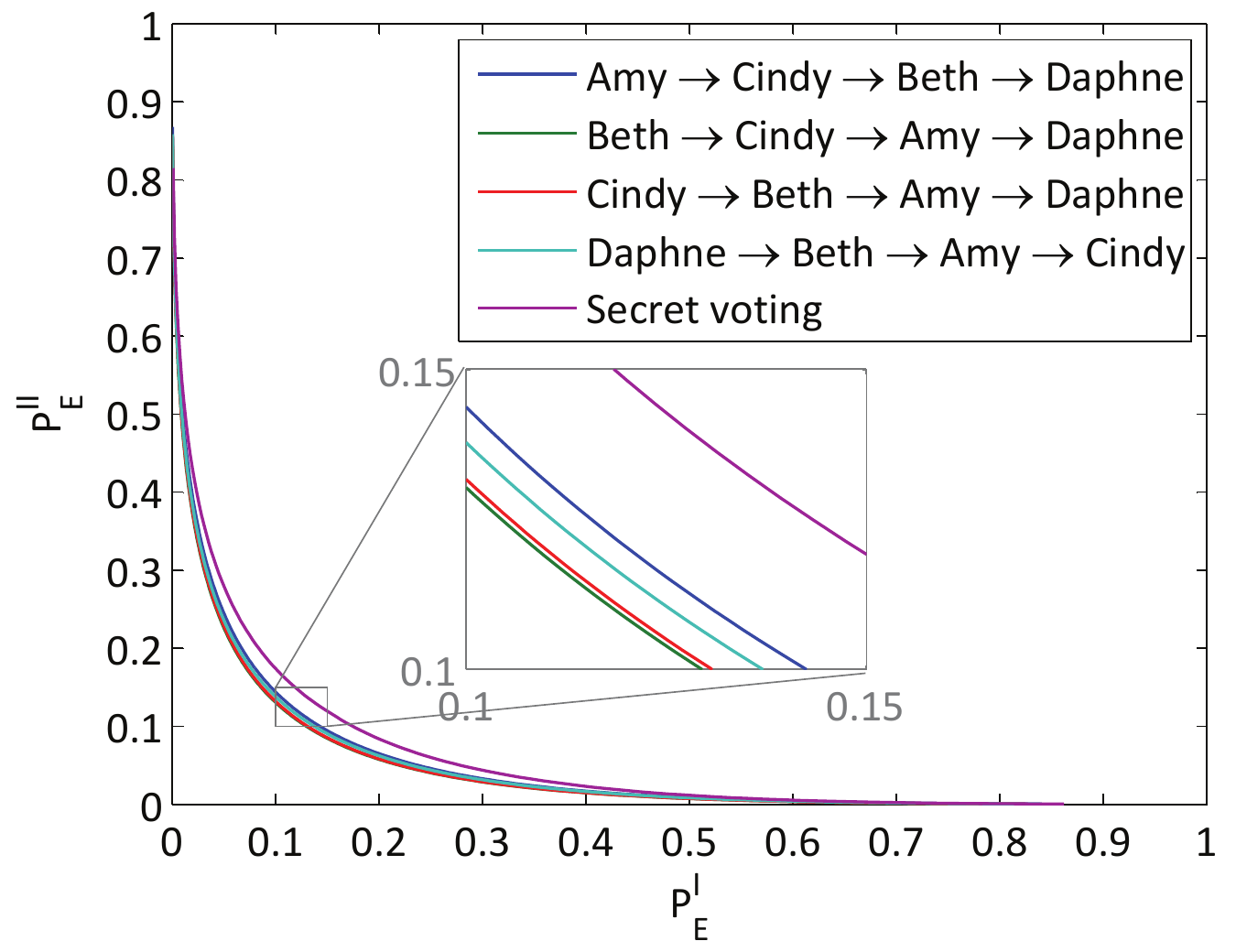}
    \caption{Lower bounds of operating regions for different orderings of four agents, Amy ($\sigma^2 = 0.25$), Beth ($\sigma^2 = 0.5$), Cindy ($\sigma^2 = 1$), and Daphne ($\sigma^2 = 2.25$), and the $2$-out-of-$4$ fusion rule.}
    \label{fig:ROC_2outof4}
\end{figure}

Figs.~\ref{fig:ROC_2outof3} and~\ref{fig:ROC_2outof4} present quantitative evidence of helpful social learning.
In these examples, we consider additive Gaussian noises with zero mean and variances $\sigma_n^2$.
The noises are mutually independent but not identically distributed.

In the case of Fig.~\ref{fig:ROC_2outof3}, there are three agents using the \textsc{majority} fusion rule.
Amy has the highest SNR ($\sigma_{\rm A}^2 = 0.25$), Beth has the median SNR ($\sigma_{\rm B}^2 = 1$), and Cindy has the lowest SNR ($\sigma_{\rm C}^2 = 2.25$).
Fig.~\ref{fig:ROC_2outof3} depicts the optimal reversed receiver operating characteristic (ROC) curves for all possible orderings of the actions of the three agents.\footnote{As decision threshold parameters are varied with the order of agent actions fixed, some set of $(\GlobalI,\GlobalII)$ pairs is achievable.  We call the lower boundary of this set the reversed ROC curve.}
We only need to consider three orderings because the order of the last two agents does not matter as we discussed in Section~\ref{sec:Experts-Crowd_AND/OR}.  The team's updated fusion rule after Amy makes a decision will be the \textsc{and} or the \textsc{or} rule, depending on Amy's decision.

There are two notable things in Fig.~\ref{fig:ROC_2outof3}.  First, the reversed ROC curve of secret voting is above that of public voting for any ordering.  This implies that public voting strictly outperforms secret voting, regardless of the order in which the agents make decisions.  Second, among the public voting scenarios, the best is when Beth makes her decision first.  Further numerical experiments for $N = 3$ with several different noise variances also show that the team performance is the best when the agent with the median SNR acts first.

Fig.~\ref{fig:ROC_2outof4} shows the optimal reversed ROC curves for four agents and the 2-out-of-4 fusion rule.  Again, we need not compare all 16 orderings.  After the first agent makes a decision, the updated fusion rule will be the 1-out-of-3 rule if the decision is 1, in which case the ordering of the rest of the agents does not matter (1-out-of-3 is a unanimity rule), and the 2-out-of-3 rule if the decision is 0, in which case we infer that the agent with median SNR should act next.  Thus, we only need to compare four orderings, where the first agent is varying and the median of the remaining three becomes the second agent.  Fig.~\ref{fig:ROC_2outof4} shows that the agent with the second-highest SNR should act first as well as that public voting always outperforms secret voting.

In this section, we have provided evidence that agents can exploit social learning to improve their team decision when the qualities of the private signals vary.  The sequence of the agents also needs to be carefully chosen to achieve the best team performance.

\section{Conclusion}
\label{sec:Conclusion}

\subsection{Summary and Interpretation of Results}

We have discussed the utility of social learning in a distributed detection and fusion system by a team of agents who have the same decision-making costs.  We showed that it is not a trivial question whether intra-team social learning is useful.  The model is intentionally designed to be simple so that we can understand fundamentals: agents have no conflicts of interests and their fusion rule is fixed and symmetric.

When Agent~$i$ casts a vote $\Hhat_i$ as an opinion on the value of the hypothesis $H$, except in degenerate situations, this vote is not independent of $H$.  Thus, when the vote is available to other agents as a public signal, it is natural for them update their beliefs about $H$; this is the basis of the social learning concept.  This paper highlights a countervailing phenomenon that arises when opinions are aggregated by voting:  the running tally of previous votes changes the effective fusion rule for the agents who have yet to act.  This fusion rule update is a less frequently explored topic.

When the agents observe conditionally iid private signals, the effects of the belief update and the evolution of the fusion rule cancel exactly.
Consequently, the optimal performance with or without public signals is the same; internally flowing information does not improve the team performance.

When the agents observe signals that are conditionally independent but not identically distributed, the existence of public signals may improve team performance, and when it does, the degree of improvement depends on the order in which the agents act.
We first showed that the $1$-out-of-$N$ and $N$-out-of-$N$ fusion rules are peculiar in that they essentially require unanimity among agents:
under the $1$-out-of-$N$ rule, all agents must vote $0$ for the team decision to be $0$;
under the $N$-out-of-$N$ rule, all agents must vote $1$ for the team decision to be $1$.
The unanimity rules do not allow a chance for the agents to specialize their decision thresholds for various public signals.
Social learning can play a role to improve team decisions as long as the fusion rule is not one of these unanimity rules.

In examples with Gaussian likelihoods, we showed that team performance improves when agents with differing observation SNRs do social learning.
With three agents making a team decision by the \textsc{majority} fusion rule, it is best for the agent with median SNR to act first.
With four agents using the 2-out-of-4 fusion rule, it is best for the agent with second-best SNR to act first.

None of our results show that the existence of public signals makes team performance worse; logically, if being influenced by public signals would be detrimental, optimal agents would simply ignore them.  We speculate, however, that public signals may be detrimental because, in human agents, belief updates overwhelm the decision rule update.  In fact, without it being explained to them, people may not have intuition for the idea that \emph{a good agent does not take lightly the disenfranchisement of later-acting agents}.

\subsection{Limitations and Possible Extensions}

Within the scope of our model, an intriguing open problem is to explain the observed patterns on the optimal order for agents with differing observation SNRs to act.  The optimal ordering is nontrivial.
Our experimentation with Gaussian likelihoods and up to four agents was extensive, but we do not know if the results extend to other likelihoods.
Are there systematic ordering rules for larger $N$?

Our model has several limitations, such as conditional independence of private signals, knowledge of the prior probability, and the agents having no motivation beyond minimization of the Bayes risk (a team performance criterion).  Each of these merits further study.
Perhaps the most interesting extensions are to the set of fusion rules.

We have assigned an equal weight to each vote in~\eqref{eq:SymmetricFusionRule} and hence called the fusion symmetric.
Suppose the fusion rule is subject to optimization.
Then one could use the Chair--Varshney fusion rule~\cite{ChairVarshney1986}, which is an algorithm that weights each individual vote according to its quality and accumulates them to compare to an optimally determined threshold.
The effect of social learning in this scenario is an open question.

Our focus on equal weighting was motivated by its prevalence in most situations in which human opinions are aggregated by voting, ranging from government elections to juries to decision-making by committees.
In human affairs, unequal weighting of votes is unusual, except for the extreme cases of veto power and a single ultimate decision maker.

Having a single ultimate decision maker changes the situation rather dramatically.
With respect to Fig.~\ref{fig:Model_N}, one may consider Alexis, Britta, \ldots, Mary to be \emph{advisers} to the \emph{decider} Norah,
and simplify the fusion rule to $\Hhat = \Hhat_N$.\footnote{This scenario gives the decider a private signal; if only advisers have private signals, then Norah could be another adviser, and the fusion center could be considered the decider.}
Here the availability of public signals does improve team performance.
This is similar to the situation studied in~\cite{RhimG2013b}, where it is shown that an adviser with an inaccurate prior probability for $H$ may provide a more helpful public signal to the decider.



\begin{thebibliography}{10}
\providecommand{\url}[1]{#1}
\csname url@samestyle\endcsname
\providecommand{\newblock}{\relax}
\providecommand{\bibinfo}[2]{#2}
\providecommand{\BIBentrySTDinterwordspacing}{\spaceskip=0pt\relax}
\providecommand{\BIBentryALTinterwordstretchfactor}{4}
\providecommand{\BIBentryALTinterwordspacing}{\spaceskip=\fontdimen2\font plus
\BIBentryALTinterwordstretchfactor\fontdimen3\font minus
  \fontdimen4\font\relax}
\providecommand{\BIBforeignlanguage}[2]{{%
\expandafter\ifx\csname l@#1\endcsname\relax
\typeout{** WARNING: IEEEtran.bst: No hyphenation pattern has been}%
\typeout{** loaded for the language `#1'. Using the pattern for}%
\typeout{** the default language instead.}%
\else
\language=\csname l@#1\endcsname
\fi
#2}}
\providecommand{\BIBdecl}{\relax}
\BIBdecl

\bibitem{Tsitsiklis1988}
J.~N. Tsitsiklis, ``Decentralized detection by a large number of sensors,''
  \emph{Math. Control, Signals, Syst.}, vol.~1, no.~2, pp. 167--182, 1988.

\bibitem{Varshney97}
P.~K. Varshney, \emph{Distributed Detection and Data Fusion}.\hskip 1em plus
  0.5em minus 0.4em\relax Springer-Verlag, 1997.

\bibitem{Banerjee1992}
A.~V. Banerjee, ``A simple model of herd behavior,'' \emph{Quart. J. Econ.},
  vol. 107, no.~3, pp. 797--817, Aug. 1992.

\bibitem{BikhchandaniHW1992}
S.~Bikhchandani, D.~Hirshleifer, and I.~Welch, ``A theory of fads, fashion,
  custom, and cultural change as informational cascades,'' \emph{J. Polit.
  Econ.}, vol. 100, no.~5, pp. 992--1026, Oct. 1992.

\bibitem{SmithS2000}
L.~Smith and P.~S{\o}rensen, ``Pathological outcomes of observational
  learning,'' \emph{Econometrica}, vol.~68, no.~2, pp. 371--398, Mar. 2000.

\bibitem{SwetsTB1961}
J.~A. Swets, W.~P. Tanner, Jr., and T.~G. Birdsall, ``Decision processes in
  perception,'' \emph{Psychol. Rev.}, vol.~68, no.~5, pp. 301--340, Sep. 1961.

\bibitem{Viscusi1985}
W.~K. Viscusi, ``Are individuals {B}ayesian decision makers?'' \emph{Am. Econ.
  Rev.}, vol.~75, no.~2, pp. 381--3853, May 1985.

\bibitem{BraseCT1998}
G.~L. Brase, L.~Cosmides, and J.~Tooby, ``Individuation, counting, and
  statistical inference: The role of frequency and whole-object representations
  in judgment under uncertainty,'' \emph{J. Experiment. Psychol.: Gen.}, vol.
  127, no.~1, pp. 3--21, Mar. 1998.

\bibitem{GillSabinSchmid2005}
C.~J. Gill, L.~Sabin, and C.~H. Schmid, ``Why clinicians are natural
  {B}ayesians,'' \emph{Brit. Med. J.}, vol. 330, pp. 1080--1083, May 2005.

\bibitem{GlanzerHM2009}
M.~Glanzer, A.~Hilford, and L.~T. Maloney, ``Likelihood ratio decisions in
  memory: Three implied regularities,'' \emph{Psychonom. Bull. Rev.}, vol.~16,
  no.~3, pp. 431--455, Jun. 2009.

\bibitem{RhimG2013b}
J.~B. Rhim and V.~K. Goyal, ``Social teaching: {B}eing informative vs. being
  right in sequential decision making,'' in \emph{Proc. IEEE Int. Symp. Inform.
  Theory}, Istanbul, Turkey, Jul. 2013.

\bibitem{ChamleySL2013}
C.~Chamley, A.~Scaglione, and L.~Li, ``Models for the diffusion of beliefs in
  social networks,'' \emph{{IEEE} Signal Process. Mag.}, vol.~30, no.~3, pp.
  16--29, May 2013.

\bibitem{OttavianiSorensen2001}
M.~Ottaviani and P.~S{\o}rensen, ``Information aggregation in debate: {W}ho
  should speak first?'' \emph{J. Publ. Econ.}, vol.~81, no.~3, pp. 393--421,
  Sep. 2001.

\bibitem{KrishnamurthyPoor2013}
V.~Krishnamurthy and H.~V. Poor, ``Social learning and bayesian games in
  multiagent signal processing: {H}ow do local and global decision makers
  interact?'' \emph{{IEEE} Signal Process. Mag.}, vol.~30, no.~3, pp. 43--57,
  May 2013.

\bibitem{RhimG2013a}
J.~B. Rhim and V.~K. Goyal, ``Keep ballots secret: {O}n the futility of social
  learning in decision making by voting,'' in \emph{Proc.\ IEEE Int. Conf. on
  Acoustics, Speech, and Signal Process.}, Vancouver, Canada, May 2013.

\bibitem{AcemogluDLO2011}
D.~Acemoglu, M.~A. Dahleh, I.~Lobel, and A.~Ozdaglar, ``Bayesian learning in
  social networks,'' \emph{Rev. Econ. Stud.}, vol.~78, no.~4, pp. 1201--1236,
  Mar. 2011.

\bibitem{Radner1962}
R.~Radner, ``Team decision problems,'' \emph{Ann. Math. Stat.}, vol.~33, no.~3,
  pp. 857--881, Sep. 1962.

\bibitem{NeymanPearson1933}
J.~Neyman and E.~S. Pearson, ``On the problem of the most efficient tests of
  statistical hypotheses,'' \emph{Phil. Trans. Roy. Soc. London Ser. A}, vol.
  231, pp. 289--337, Jan. 1933.

\bibitem{Tsitsiklis1993}
J.~N. Tsitsiklis, ``Decentralized detection,'' in \emph{Advances in Statistical
  Signal Processing}, H.~V. Poor and J.~B. Thomas, Eds.\hskip 1em plus 0.5em
  minus 0.4em\relax Greenwich, CT: JAI Press, 1993, pp. 297--344.

\bibitem{RhimVG2012c}
J.~B. Rhim, L.~R. Varshney, and V.~K. Goyal, ``Quantization of prior
  probabilities for collaborative distributed hypothesis testing,''
  \emph{{IEEE} Trans. Signal Process.}, vol.~60, no.~9, pp. 4537--4550, Sep.
  2012.

\bibitem{ChairVarshney1986}
Z.~Chair and P.~K. Varshney, ``Optimal data fusion in multiple sensor detection
  systems,'' \emph{{IEEE} Trans. Aerosp. Electron. Syst.}, vol. ASE-22, no.~1,
  pp. 98--101, Jan. 1986.

\end{thebibliography}
\end{document}